\documentclass{article}
\usepackage[utf8]{inputenc}
\usepackage[colorlinks=true, allcolors=blue]{hyperref}
\usepackage{graphicx}
\usepackage{caption}
\usepackage{url}
\usepackage{enumerate}
\usepackage{csquotes}   
\usepackage{longtable}  
\usepackage{hang}
\usepackage{amsmath}
\usepackage{epigraph}   
\usepackage{soul}       
\usepackage{subfig}
\usepackage{natbib}
\providecommand{\keywords}[1]
{
  \small	
  \textbf{\textit{Keywords---}} #1
}

\usepackage{lineno}

\title{Impact of the 2022 OSTP Memo: A Bibliometric Analysis of U.S. Federally Funded Publications, 2017-2021}

\author{Eric Schares\\Iowa State University\\\url{eschares@iastate.edu}\\ORCiD 0000-0002-6292-8221}
\date{}

\begin{document}
\maketitle

\begin{abstract}

On August 25, 2022, the White House Office of Science and Technology Policy (OSTP) released a memo regarding public access to scientific research.
Signed by Director Alondra Nelson, this updated guidance eliminated the 12-month embargo period on publications arising from U.S. federal funding that had been allowed from a previous 2013 OSTP memo.


While reactions to this updated federal guidance have been plentiful, to date there has not been a detailed analysis of the publications which would fall under this new framework. 
The OSTP released a companion report along with the memo, but it only provided a broad estimate of total numbers affected per year.

Therefore, this study seeks to more deeply investigate the characteristics of U.S. federally funded research over a 5-year period from 2017-2021 to better understand the updated guidance's impact. It uses a manually created custom filter in the Dimensions database to return only publications that arise from U.S. federal funding.

Results show that an average of 265,000 articles were published each year that acknowledge U.S. federal funding agencies, and these research outputs are further examined by publisher, journal title, institutions, and Open Access status.

Interactive versions of the graphs are available online at \url{https://ostp.lib.iastate.edu/}.

\end{abstract}

\vspace{1em}
\keywords{OSTP memo, Open Access, federally funded research, public access, embargo, Dimensions}

\section{Introduction}

\epigraph{OSTP Issues Guidance to Make Federally Funded Research Freely Available Without Delay}{\textit{\citeauthor{WH_press_release}}}

On August 25, 2022, the White House Office of Science and Technology Policy (OSTP), under Director Alondra Nelson, released new policy guidance
entitled ``Ensuring Free, Immediate, and Equitable Access to Federally Funded Research" \citep{Nelson_memo}.
The memo states that, by 2026:
\begin{displayquote}
    ``all peer-reviewed scholarly publications authored or coauthored by individuals or institutions resulting from federally funded research \textbf{are made freely available and publicly accessible by default in agency-designated repositories without any embargo or delay after publication}" [emphasis preserved from original]
\end{displayquote}

This new policy framework is an update to previous guidance on public access to scientific research.
A 2013 policy released by Director John Holdren allowed a 12-month embargo on publications arising from federal funding, and only applied to federal agencies which granted over \$100 million annually \citep{Holdren_memo}.
By contrast, the new 2022 Nelson memo eliminates the possibility of a 12-month embargo period for federally funded peer-reviewed research articles that was allowed under the previous 2013 guidance.
It also extends guidance to the data underlying those publications, strengthens the data sharing plans, contains specific metadata requirements,
and applies to \emph{all} U.S. federal granting agencies regardless of their annual granting amounts
\citep{breakthroughs_for_all_pressrelease}.
The Association of Research Libraries has released a summary table outlining the details of the 2013 and 2022 OSTP memos for ease of comparison \citep{ARL_comparison_table}.

Reactions to the Nelson memo have been plentiful and varied, with
the release garnering national news coverage \citep{NYTimes}.
Libraries \citep{ARL},
universities \citep{AAU},
librarians \citep{Cambridge, SKitch_initial},
societies \citep{SPARC, Coalition_S},
consultants \citep{Clarke_and_Esposito, DeltaThink_newsandviews_OSTP},
publishers \citep{AssnAmPubl, PLOS, IOP}, 
funders \citep{Tananbaum_OpenFunders},
and researchers \citep{Am_Anthro_Assn} 
have weighed in with statements or opinion pieces, some more enthusiastic about the development than others.

Chairwoman of the House Committee on Science, Space, and Technology Eddie Bernice Johnson and Ranking Member Frank Lucas sent a joint letter to
the newly confirmed Director of the OSTP, Dr. Arati Prabhakar asking for clarifications \citep{Johnson_OSTP_questions}. 
Questions still remain about the 2026 implementation and how specific practices will result from this new guidance, how agencies will update their policies, and concerns about participation in research if article processing charges (APCs) are increased or used more widely.






\section{Research Questions}
\label{subsec:RQ_list}

In addition to the 2022 memo, the OSTP also released a companion report on the potential economic impact of the updated guidance and its effects on federal grant funding agency policies \citep{OSTP_economic_impact}.
The report estimates ``between 195,000 and 263,000 articles were federally funded in 2020" but does not provide a more granular breakdown of these articles.
An additional analysis estimates 197,000 federally funded articles in 2021 \citep{SK_Petrou}.
Other than these high-level studies,
there have been limited analyses to more fully detail the characteristics of publications that fall within this newly expanded scope.

Therefore, this study seeks to address the following research questions:

\begin{labeledlist}{RQ3}
\item[\textbf{RQ1}]     How many U.S. federally funded publications have there been over the past five years? What are the yearly totals, and what proportion do these represent of worldwide and United States-specific output?

\item[\textbf{RQ2}] Which U.S. federal funding agencies awarded these grants?

\item[\textbf{RQ3}] How do the number of federally funded articles vary by research category/discipline?

\item[\textbf{RQ4}] Which publishers tend to publish federally funded articles?

\item[\textbf{RQ5}] Which journals tend to publish federally funded articles?

\item[\textbf{RQ6}] Which research institutions are authors who tend to publish federally funded articles affiliated with?

\item[\textbf{RQ7}] In what manner were these federally funded articles published? Were they published openly or behind a paywall?

\end{labeledlist}

\section{Methodology}
\label{sec:Methodology}

The analysis was conducted using the bibliometric database Dimensions, available at \url{http://app.dimensions.ai}.
This study used the paid version of the tool;
there is also a free version available, though with limited functionality.
Dimensions ingests metadata from Crossref to make connections across publications, authors, funders, institutions, and more \citep{QSS_Dim}. 
The database uses this as a starting point and further enriches funding information by analyzing text provided in authors' acknowledgments sections and through agreements with publishers to to obtain additional funding information.

Dimensions was chosen for this study because of relevant advantages over other commonly used bibliometric databases.
It indexes a wider range of journals and has more complete coverage than Web of Science, which is estimated to cover only 10-12\% of journals \citep{Clarivate_WoS_coverage}.
OpenAlex, a free and open bibliographic database, also uses metadata reported to Crossref by publishers, as well as data from the now-discontinued Microsoft Academic Graph, scraping publisher websites, and other sources \citep{OpenAlex_arXiv_STI}.
However, OpenAlex does not include funding information in its records of works.

This study is particularly affected by funding information that is deposited to Crossref and included in Dimensions.
The availability of major metadata elements in Crossref was quantified by \citet{Waltman_Crossref_funderinfo}, who found 25\% of articles in 2020 reported some funding information.
\citet{Kramer_QSS_fundingmetadata} specifically analyzed funder information in several bibliometric data sources and quantified the extraction of additional funding information from acknowledgment text, going beyond what is deposited by publishers to Crossref. 
Web of Science, Scopus, and Dimensions all infer this additional funding information. 
Dimensions reported funding information on 81\% of the records in a study of publications by the Dutch Research Council, compared to 67\% availability in Crossref. However, the information was inconsistent, with not all publications correctly naming the funder or providing the funder ID. The performance also varied considerably by publisher.

A case study deeply analyzing one example funding statement clearly illustrates the difficulties in untangling personal, financial, and logistical acknowledgments in the same section. Different bibliometric databases and tools are also shown to have different interpretations of the same funding information \citep{LeidenMadtrics_casestudyfundingack}.


\subsection{Two Possible Approaches}
The most crucial part of this analysis was defining the custom Funder group, which controls which publications are included/excluded in the analysis. The Dimensions web interface offers the ability to create a custom group of any single facet type; in this case, Funders.
There were two main options to consider when deciding how to construct this custom group - define what to \emph{include}, or define what to \emph{remove}.

Attempting to include all federal grant-funding agencies in one custom Funder Group was the first attempt. In theory, this sounds like the simpler approach; only keep those agencies whose funded output would qualify under the new OSTP guidance. 
Additionally, the OSTP economic impact report states that just six federal agencies ``account for more than 94 percent of the approximately \$150 billion" in federal research and development \citep{OSTP_economic_impact}.
This means the filter would be very nearly complete after including only six agencies.

However, 
it quickly became apparent that identifying and building one custom filter that covered all possible agencies, divisions, institutes, centers, and their name variants was not feasible. 
For example, the U.S. Department of Health and Human Services (HHS) is a large federal-granting agency. Within it are several Operating Divisions, such as the National Institute for Health (NIH) or the Food and Drug Administration (FDA).
Within each of these Operating Divisions are further institutes and centers, such as the NIH's National Institute on Aging or the Office of AIDS Research. 
Dimensions enriches the publisher-supplied metadata from Crossref with additional information from a publication's acknowledgments section, but authors do not consistently identify funder names. Dimensions takes what it can find and does not further correlate or gather these variants into coherent groups.
Depending on what is specifically acknowledged in a publication, the funding information returned may be as granular as a specific division, or as broad as an entire agency.

With this approach considered unmanageable, work then turned to the second option which channeled the sculptor Michelangelo: remove everything that is not a federal funder.
However, this too had a fatal flaw. 
While the funder information was more consistent and the granular nature of federal data was not a concern, removing private foundations, 501(c)(3)'s, corporations, nonprofits, state agencies, or other organizations had the unintended effect of also removing desired publications.
If an article acknowledged funding from both a foundation and a federal agency, the fact that the foundation was being removed from the analysis meant the entire paper would be excluded, even though it did have federal funding and should rightly be included in the dataset.

Therefore, the final answer turned out to be a combination of the two approaches.

\subsection{Defining Custom Funder Group}
With the Dimensions web interface limited to the years 2017-2021 and Country of Funder set to United States, the Analytical View for Funders was able to quickly export the top 500 funders to meet those criteria (both federal and non-).

Once the funder names were exported to an Excel sheet, some funders from countries other than the U.S. were still present due to publications with support from multiple grants and international collaborations.
Limiting the exported column Country to United States reduced the number of funders from 500 to 331.
It was then a manual task to search each funder one a time, investigate its status, and determine if it was a federal agency or not.
Those which were found to be foundations, 501(c)(3)'s, corporations, nonprofits, state agencies, or other organizations were flagged.
Each of these non-federal funders was then added to a second, temporary custom Funder Group in Dimensions in order to exclude them all in one large batch.

In the first round of investigation, 193 of the 331 organizations (58\%) were determined to be non-federal funders,
and the process of exporting, filtering, and manual investigation was repeated two more times. 
This uncovered 87 and 36 more funding agencies to remove, respectively, for a total of 316 non-federal funders.
The process was halted there, since the non-federal funders were showing relatively low publication activity ($<156$ over 5 years, or approx. 30 publications per year) and there were diminishing returns when going further.

With the 316 non-federal funders excluded, there were 1.129M publications remaining. However, this is not the number to be concerned with, since by definition it will be an under-count.
Some papers with federal funding get excluded by the fact that they \emph{also} have non-federal funding.
Now that non-federal agencies are cleaned out, the list of funders exported from the Analytical view has only federal granting agencies remaining, giving a cleaner picture of the dispersed and fractured naming conventions.
These 177 federal agencies were then added to a custom filter one by one, with the number of newly qualifying papers recorded after each agency was added.

The last step was to look at the existing Funder Groups in Dimensions that appear to all users.
These pre-defined groups attempt to reconcile the fractured naming conventions for a few federal agencies, including CDC, DoD, DoE, NIH, NOAA, NSF, NASA, and USDA. 
These existing groups were expanded and the individual centers within were compared to the 175 federal agencies that made up the new custom group thus far. 
62 additional funders were found and added to the list, for a total of 239 federal funding agencies. 
We can be reasonably confident in the completeness of this final custom group. By the end of the list, agencies are typically contributing single digits worth of unique publications to the whole, or a thousandth of a percent.

The full list of funders that make up the custom Funder Group of U.S. federal granting agencies is included in the Appendix.
\section{Results}

The Dimensions web search interface was used for the analysis. The Dimensions API was considered and investigated briefly, but it would take a non-trivial amount of work to process text fields and strings of funder information. The web interface also offers a wide range of pre-built Analytical Views, which were very useful when conducting this analysis.

Years of publication were limited to 2017, 2018, 2019, 2020, or 2021.
The Publication Type ``Preprint” was excluded from the analysis, as the OSTP memo states only peer-reviewed publications qualify under its guidance. Preprints are versions of publications before peer review has been conducted; however, it is worth noting that several recent initiatives have begun to offer or organize peer reviews of preprints \citep{PeerCommunityIn, ReviewCommons, PREreview}.
All other document types, including Article, Proceeding, Chapter, Edited Book, and Monograph, were included since
the memo states that it applies to ``scholarly publications."
No restrictions or distinctions were made between a corresponding author on a publication and a researcher's name appearing anywhere in the author list.

\begin{figure}[h!] 
    \centering 
    \includegraphics[width=.75\textwidth]{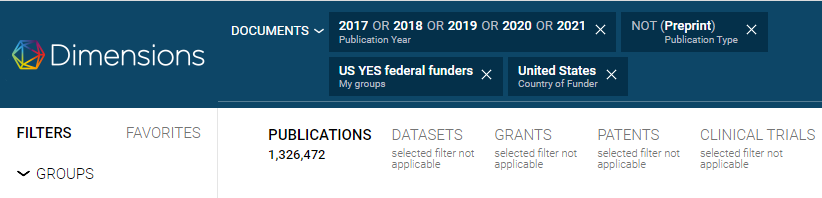} 
    \caption{Dimensions search query}
    \label{fig:Dim_search_query} 
\end{figure}

The full search run in Dimensions on October 18, 2022 is shown in Figure \ref{fig:Dim_search_query}:
\begin{itemize}
    \item Publication Year = 2017-2021
    \item Publication Type = NOT Preprint
    \item Country of Funder = United States
    \item Funder = Custom U.S. federal funders group
\end{itemize}

With the dataset defined, we can now move on to answering the specific research questions outlined in Section \ref{subsec:RQ_list}.

\subsection{RQ1 - How many federally funded (FF) publications?}
\label{Results_subsec:RQ1}
The search resulted in a total of 1,326,472 publications. In general, approximately 250,000 - 277,000 publications per year are a result of U.S. federal funding (Table \ref{tab:totalnumber_pubs_by_year}) and would thus be affected under the new OSTP guidance.
The 5-year average is 265,294 publications.
This falls on the high end of the range quoted by the OSTP economic impact memo of 195k-263k annually \citep{OSTP_economic_impact}.

Comparing just the year 2021, the result is nearly 40\% higher than the 197,000 U.S. federally funded articles estimated by an analysis using the database Web of Science \citep{SK_Petrou}. As indicated above, Dimensions indexes more broadly and captures a wider range of journals and their articles.

\begin{table}[h!]
    \centering
    \caption{Number of U.S. federally funded publications by year}
    \label{tab:totalnumber_pubs_by_year}
    \begin{tabular}{|c|c|}
    \hline
    Year & \# of publicaions \\ \hline
    2021 & 275,825           \\ \hline
    2020 & 277,407           \\ \hline
    2019 & 262,682           \\ \hline
    2018 & 259,518           \\ \hline
    2017 & 251,040           \\ \hline \hline
    Total & 1,326,472       \\ \hline
\end{tabular}
\end{table}

The total of 1.3M represents
33\% of all United States domestic research output over these five years (n=4,020,840), defined as any publication type, funded or non-, with at least one author from a U.S. institution (Location - Research Organization - Country/Territory = United States).
This matches the 31\% federally funded output found by \citet{SK_Petrou}.
It also represents 4.47\% of total global research output over those five years (n=29,646,485).

\subsection{RQ2 - Which Federal granting agencies?}
\label{Results_subsec:RQ2}

Table \ref{tab:top10funderstable} shows the top 10 U.S. federal granting agencies in terms of number of resulting publications over the 5-year period in this study. It is important to note that these funding agency names appear as they come from Dimensions, and are not synthesized or otherwise combined. This is most readily apparent when looking at the multiple NSF agencies that appear in the top 10: NSF MPS, NSF CISE, and NSF ENG.
Beyond this top 10, a complete list of 319 grant making agencies, both U.S. federal and non-, and their number of related publications is provided with the data availability statement.

\begin{table}[h!]
  \centering
  \caption{Top 10 U.S. funding agencies by number of resulting publications}
    \begin{tabular}{|l|c|}
    Name  & \multicolumn{1}{l}{\# Publications} \\ \hline
    National Cancer Institute (NCI) & 137,496 \\
    Directorate for Mathematical \& Physical Sciences (NSF MPS) & 118,881 \\
    National Institute of General Medical Sciences (NIGMS) & 118,095 \\
    United States Department of Energy (DOE) & 113,712 \\
    National Heart Lung and Blood Institute (NHLBI) & 89,767 \\
    Directorate for Comp \& Info Science \& Engr (NSF CISE) & 80,068 \\
    Directorate for Engineering (NSF ENG) & 77,784 \\
    National Center for Advancing Translational Sciences (NCATS) & 77,170 \\
    National Institute of Allergy and Infectious Diseases (NIAID) & 76,801 \\ \hline
    \end{tabular}%
  \label{tab:top10funderstable}%
\end{table}%

Exporting funder information is limited to the top 500 funders at a time in the Dimensions web interface. It would seem possible to export, then sum the number of publications from each funder to see how many of the total 1.326M results are included in the top 500 (``Publications" column of Table \ref{tab:top10funderstable}).
However, while the web interface de-duplicates results and only returns the actual count of unique publications, the exported data \emph{is not} de-duplicated.
Each agency is listed with the total number of papers that acknowledge funding from it.
This means that a single paper that includes funding information from two or more granting agencies would be counted two or more times in the exported file, one for each named agency.
As a result, adding up the counts from the Funder export file gives a sum of 2.58M records.

The large number of qualifying federal funders in the custom filter is one of the major changes in the 2022 OSTP memo, as the 2013 version applied only to those U.S. federal granting agencies which award more than \$100M annually.
For example, the United States Department of Agriculture (USDA)'s Animal and Plant Health Inspection Service (APHIS) award total for FY 2020 was \$19M \citep{NSF_funding_tables}, so it was not covered under the 2013 Holdren memo.
USDA APHIS was acknowledged on 1,693 publications over the five year period studied here, coming in at rank \#149. 
In 2021, 107 of the 399 articles acknowledging funding from APHIS were published behind a paywall (26\%).
The updated Nelson memo guidance does away with the \$100M annual funding threshold, so APHIS will start to be covered by the zero embargo policy when it goes into effect by 2026.

\subsection{RQ3 - Which research categories are FF publications in?}
\label{Results_subsec:RQ3}
The number of publications affected by the updated OSTP memo will vary considerably depending on research field. Some disciplines are more publication intensive, and some rely more heavily on federal grant monies.

\begin{figure}[h!t] 
\centering 
  \includegraphics[width=.75\textwidth]{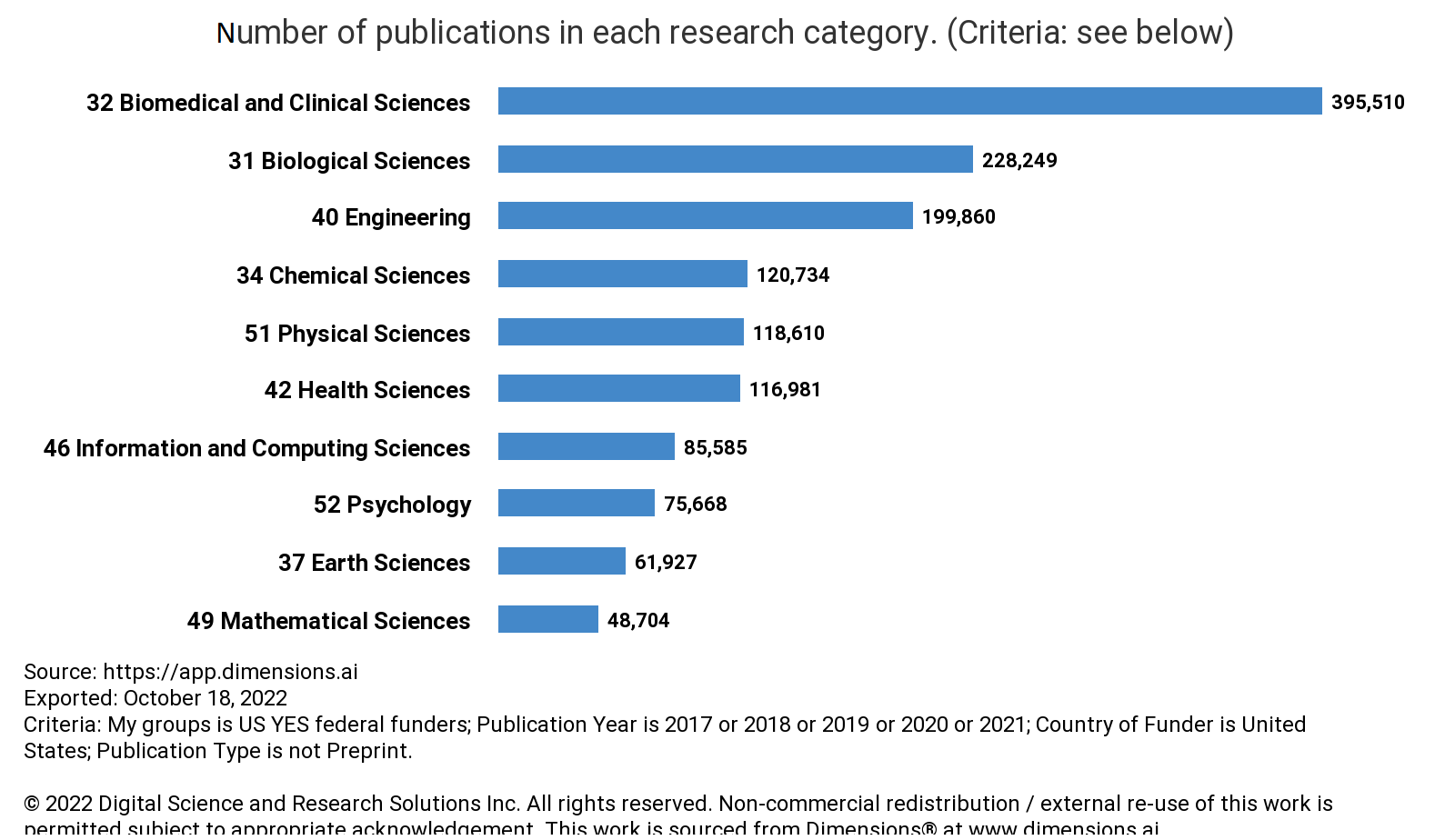} 
  \caption{Research Category classifications of publications, 2017-2021}
  \label{fig:research_categories} 
\end{figure} 

Figure \ref{fig:research_categories} shows the 10 highest research categories in terms of number of publications, taken from a Dimensions Analytical view in the web interface.
Dimensions uses the ANZSRC 2020 category classifications for discipline coding \citep{ANZSRC_codes}.
Looking at the top three categories, Biomedical and Clinical Sciences published nearly 400,000 total articles over these five years.
Biological Sciences and Engineering were numbers 2 and 3 at around 200,000 publications each, and groups 4-6 each published around 120,000 total articles.


\subsection{RQ4 - Which publishers tend to publish FF articles?}
\label{Results_subsec:RQ4}

\begin{figure}[h!t] 
\centering 
  \includegraphics[width=1\textwidth]{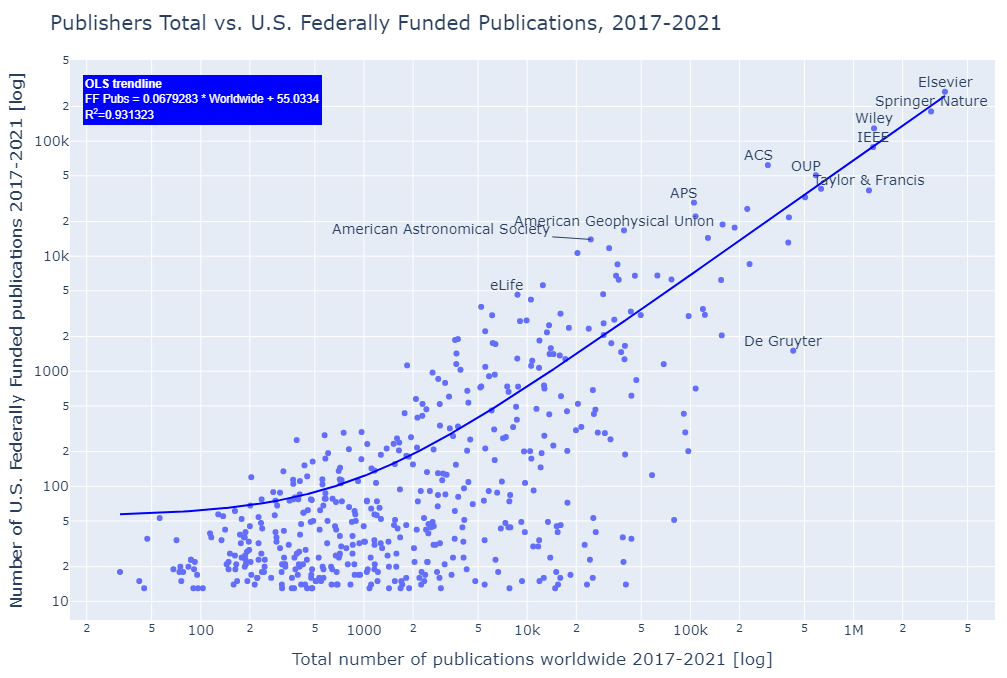} 
  \caption{By Publisher - U.S. federally funded publications vs. Total publications. Interactive version at \url{https://ostp.lib.iastate.edu/}}
  \label{fig:publishers_FF_vs_worldwide} 
\end{figure}

Publishers vary in the subject matter they produce, so the impact of the new OSTP guidance will affect publishers to different degrees.
Figure \ref{fig:publishers_FF_vs_worldwide} shows the top 500 publishers in terms of number of U.S. federally funded outputs. It plots the number of FF publications from 2017-2021 vs. the total number of publications over those years, broken out by publisher.
The trendline shows a strong relationship between the two types; as the total number of publications increases, so too does the number of federally funded articles tend to increase.

Some interesting data points begin to appear. 
Wiley, American Chemical Society (ACS), American Physical Society (APS), American Geophysical Union (AGU), the American Astronomical Society, eLife, and others all appear as points above the trendline. 
This means they publish more federally funded research than might be expected from the overall trend.

Many data points are clustered in the lower left corner of Figure \ref{fig:publishers_FF_vs_worldwide}, causing data labels to overlap and preventing them all from being shown.
To facilitate deeper exploration and allow the user to investigate data points of interest, interactive versions of each plot shown in this analysis are available online at \url{https://ostp.lib.iastate.edu/}.

\begin{figure}[h!t] 
\centering 
  \includegraphics[width=1\textwidth]{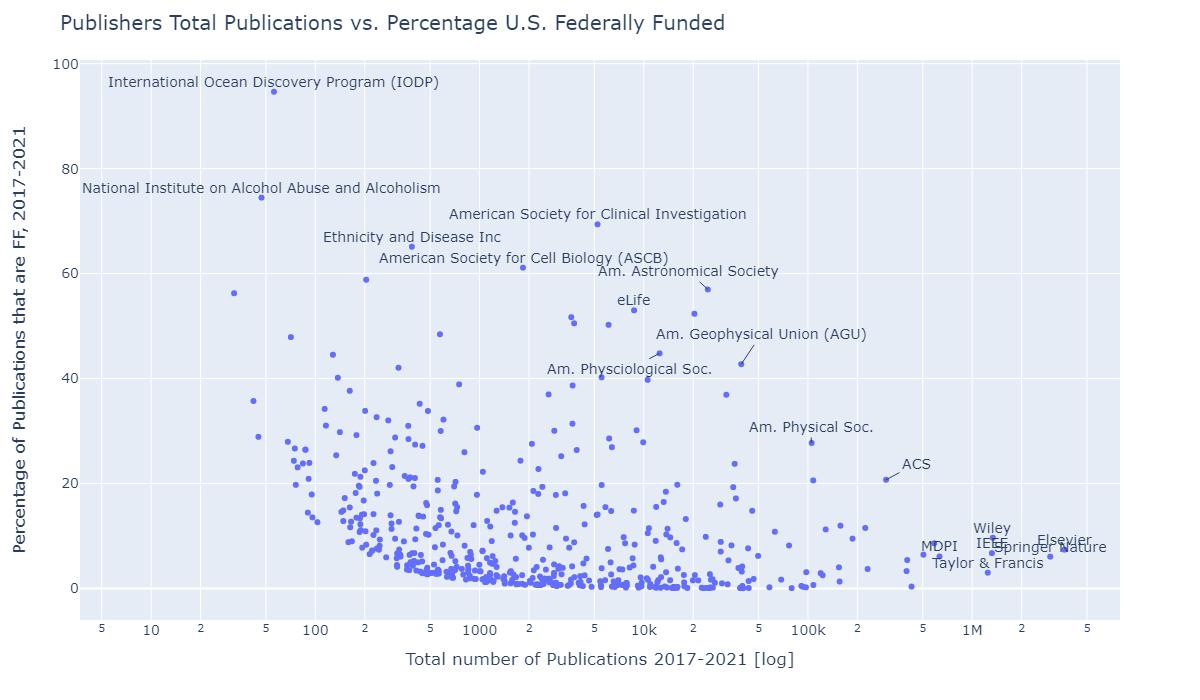} 
  \caption{By Publisher - Percentage of federally funded publications vs. Total number. Interactive version at \url{https://ostp.lib.iastate.edu/}}
  \label{fig:publishers_worldwide_vs_percent} 
\end{figure} 

\begin{table}[h!tbp]
  \centering
  \caption{Publishers with highest Percentage of FF Pubs}
    \begin{tabular}{|l|ccc|}
    Publisher  & \multicolumn{1}{l}{FF Pubs} & \multicolumn{1}{l}{All Pubs} & \multicolumn{1}{l}{\% FF} \\ \hline
    Intl Ocean Discovery Program (IODP) &  53  &  56  & 94.64 \\
    Natl Inst on Alcohol Abuse and Alcoholism &  35  &    47  & 74.47 \\
    Am Soc for Clinical Investigation &  3,627  & 5,228  & 69.38 \\
    Ethnicity and Disease Inc &  252  &  387  & 65.12 \\
    Am Society for Cell Biology (ASCB) &  1,124  &   1,839  & 61.12 \\
    Tobacco Regulatory Science Group & 120  &    204  & 58.82 \\
    American Astronomical Society &  13,980  &     24,541  & 56.97 \\
    Academy of Natl Sciences of Philadelphia &    18  &   32  & 56.25 \\
    eLife &  4,625  &   8,728  & 52.99 \\
    Proc of the Natl Academy of Sciences &    10,649  &   20,344  & 52.34 \\
    Rockefeller University Press &    1,870  &   3,618  & 51.69 \\
    The Am Assn of Immunologists &  1,903  &   3,767  & 50.52 \\
     \hline
    \end{tabular}%
  \label{tab:publishers_by_highestFFpercentage}%
\end{table}%

It may be tempting to say that publishers above the trendline will be more affected by the new guidance, as more of their papers will fall under zero-embargo with immediate public access.
However, Figure \ref{fig:publishers_FF_vs_worldwide} is showing only the absolute numbers of publications.
Further, the x- and y-axes show numbers on a log scale.
Therefore, it can be difficult to determine the exact proportion of articles that are federally funded for each publisher, since the scales are not linearly increasing.

Figure \ref{fig:publishers_worldwide_vs_percent}, then, continues the investigation by showing the \emph{percentage} of each publishers' total output that arise from U.S. federal funding.
In general, most publishers cluster between 0 and 20\%, with thinning numbers from 20-40\% and 21 out of 500 publishers above 40\% FF.
The highest percentage of FF research are among publishers that have relatively low total output.

Here again we can see the American Astronomical Society and eLife having relatively high percentages of FF articles. 
However, additional publishers jump to the top,  such as the 
International Ocean Discovery Program (95\% of total),
National Institute on Alcohol Abuse and Alcoholism (75\%),
American Society for Clinical Investigation (69\%),
and others.
Table \ref{tab:publishers_by_highestFFpercentage} shows the twelve highest publishers by federal funding percentage, though many publishers listed publish relatively few articles overall.
The American Astronomical Society has the highest total output of those with over 50\% federally funded, at 24,541 total articles in the five years studied here.

The American Physical Society (APS) 
and American Chemical Society (ACS)
move farther out on the x-axis with higher total output and
lower on the y-axis when looking at relative percentages in Figure \ref{fig:publishers_worldwide_vs_percent}, to 27.7\% and 20.7\%, respectively.
Wiley was also above the overall trendline in absolute numbers, but shows less than 10\% of its total output as a result of federal funding.

\subsection{RQ5 - Which journals tend to publish FF research?}
\label{Results_subsec:RQ5}

Next we look at the individual journal level, starting again with the top 500 titles in terms of the number of U.S. federally funded publications from 2017-2021, and plotting vs. the total number from those years.

\begin{figure}[h!t] 
\centering 
  \includegraphics[width=1\textwidth]{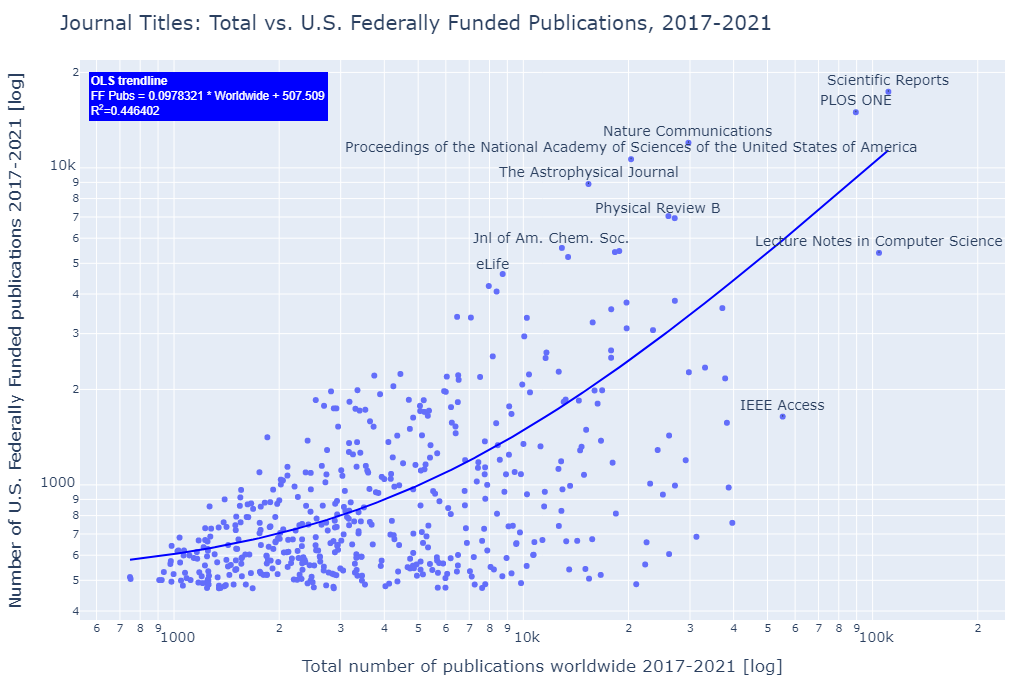} 
  \caption{By Journal - U.S. federally funded publications vs. Total publications. Interactive version at \url{https://ostp.lib.iastate.edu/}}
  \label{fig:journals_worldwide_vs_FF} 
\end{figure}

Scientific Reports and PLOS ONE both publish a very high number of federally funded research, but they also publish a very high number of articles in general (top right corner of Figure \ref{fig:journals_worldwide_vs_FF}).
IEEE Access also publishes a large number of total articles, but relatively few of them are the result of federal funding.
Nature Communications, PNAS, and the Astrophysical Journal all published between 8,000-12,000 FF articles during the time period studied here, much higher than would be expected from the trendline.
However, as with Section \ref{Results_subsec:RQ4} on Publishers, it is difficult to determine the percentage of publications with federal funding within each journal from this graph.

\begin{figure}[h!t] 
\centering 
  \includegraphics[width=1\textwidth]{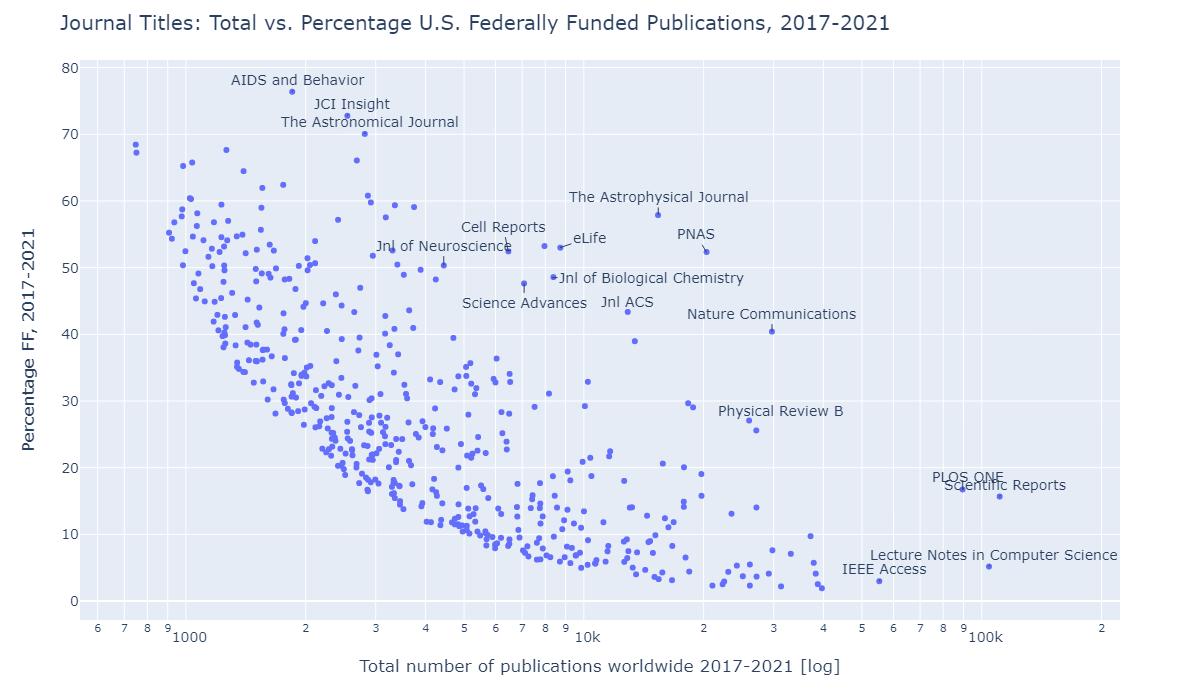} 
  \caption{By Journal - Percentage of federally funded publications vs. Total number. Interactive version at \url{https://ostp.lib.iastate.edu/}}
  \label{fig:journals_worldwide_vs_FFpercentage} 
\end{figure}

Figure \ref{fig:journals_worldwide_vs_FFpercentage} moves from absolute numbers to percentages and shows a similar trend to publishers, but with the curve more filled in.
Instead of being very scattered and vertically dispersed with large gaps between data points at high percentage levels, the journal-level is more smoothly spread out as the curve moves to the top-left.
More titles are present in the $>$40\% region of the chart, with 120 out of 500 data points falling in this range.

Nevertheless, the highest percentage of FF research at the journal level again occurs in journals which publish relatively few articles overall.
64 journals have 50\% or more of their total output from U.S. federally funded research.
AIDS and Behavior has the highest FF percentage at 76\%, followed by JCI Insight and the Astronomical Journal, with both over 70\%.
Table \ref{tab:journaltitles_by_highestFFpercentage} lists the journals with the highest FF percentages and their corresponding numbers.
When looking at percentages of FF research, the large total volume destinations Scientific Reports and PLOS ONE mentioned earlier both come in at around 16\% each, with IEEE Access at slightly less than 3\%.

\citet{SK_Petrou} looked deeply at the case of four journals in particular, some ``of the most prestigious, mostly paywalled, scholarly journals": Nature, Science, Cell, and PNAS.
The finding was that more than 40\% of these journals' papers were from U.S. federally funded research.
This analysis agrees in the cases of Cell and PNAS (60\% and 52\%), but differs for Nature and Science. 
In those cases, this study finds only 15 and 13\% of papers, respectively, to be a result of federal funding over 5 years.
One possible explanation is the extensive front matter and high level of editorial content in these journals, which is included in the Dimensions ``Article" type, but is provided as a separate facet and thus able to be filtered out of results in Web of Science (see also Section \ref{sec:Limitations}).
However, even if one were to take the Dimensions FF number as the numerator (casting the widest possible net to find FF articles) and the WoS number as the denominator (narrowing to a stricter definition of Article type), the percentages only increase modestly to 22 and 23\%.

\begin{table}[h!tbp]
  \centering
  \caption{Journals with highest percentage of FF articles}
    \begin{tabular}{|l|ccc|}
    \textbf{Journal Title} & \multicolumn{1}{l}{\textbf{FF Pubs}} & \multicolumn{1}{l}{\textbf{All Pubs}} & \multicolumn{1}{l}{\textbf{Percent FF}} \\ \hline
    
    AIDS and Behavior & 1,413  & 1,850  & 76.38 \\
    JCI Insight & 1,852  &  2,545  & 72.77 \\
    The Astronomical Journal & 1,972  &  2,815  & 70.05 \\
    Preventing Chronic Disease &  512  &  748  & 68.45 \\
    Alcoholism Clinical \& Expt'l Research &   855  &    1,264  & 67.64 \\
    Genes \& Development &   505  & 751  & 67.24 \\
    Journal of Clinical Investigation &  1,776  & 2,688  & 66.07 \\
    Contemporary Clinical Trials &  682  & 1,037  & 65.77 \\
    Jnl of Substance Abuse Treatment & 642  & 984  & 65.24 \\
    Molecular Biology of the Cell & 900  &  1,396  & 64.47 \\
    American Jnl of Preventive Medicine & 1,096  & 1,756  & 62.41 \\
    Neuropsychopharmacology & 964  & 1,556  & 61.95 \\
    
    \hline
    \end{tabular}%
  \label{tab:journaltitles_by_highestFFpercentage}%
\end{table}%

\subsection{RQ6 - Which research institutions are authors who tend to publish FF research affiliated with?}

\label{Results_subsec:RQ6}

Research institutions will also see effects to varying degrees from the updated OSTP memo. This is what will likely be of most interest to individual libraries and universities - how will our specific campus be affected by this new guidance?

The analysis in this section was filtered down to look only at research institutions in the U.S (Dimensions filter: Location - Research Organization - Country/Territory = United States).
Note that organizations from around the world will also be affected by the new OSTP memo, not only those in the United States.
When publishing research collaboratively with a researcher that receives some federal funding from the United States, the resulting research output will still qualify under the updated guidance and be made publicly available immediately.
Therefore, institutions from around the world do appear in this section, but with only a fraction of their total numbers represented.

Once again, the general trend holds: as the total number of publications goes up, so too does the number of federally funded publications.
Three institutions that stand out as publishing more funded research than typical are three national labs: Lawrence Berkeley, Oak Ridge, and Argonne. This makes sense that more of their research would be a result of federal funding as the labs themselves are the result of federal funding.

\begin{figure}[h!t] 
\centering 
  \includegraphics[width=1\textwidth]{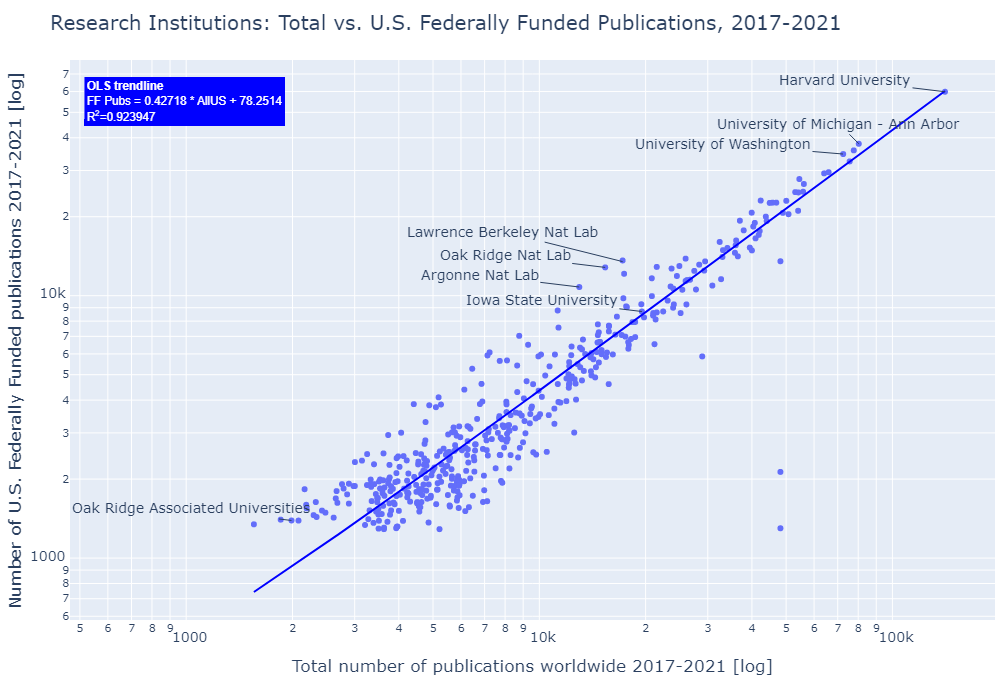}
  \caption{By Research Institution - U.S. federally funded publications vs. Total publications. Interactive version at \url{https://ostp.lib.iastate.edu/}}
  \label{fig:ResOrgs_worldwide_vs_FF} 
\end{figure} 

\begin{figure}[h!t] 
\centering 
  \includegraphics[width=1\textwidth]{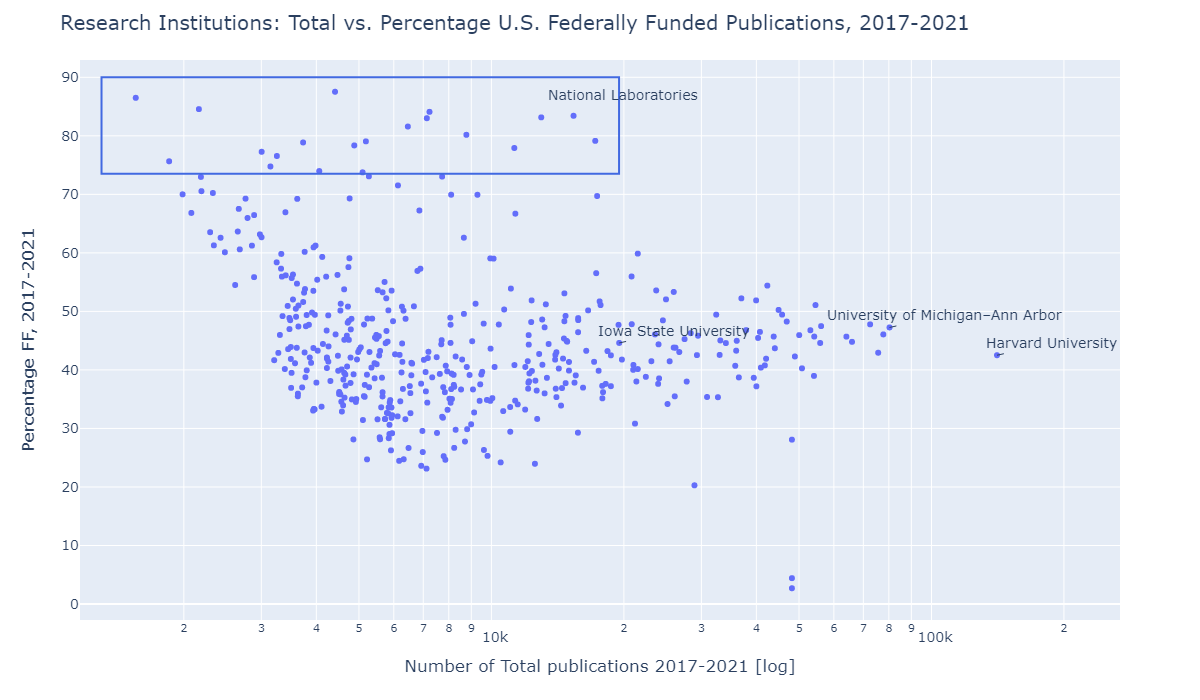} 
  \caption{By Research Institution - Percentage of federally funded publications vs. Total number. Interactive version at \url{https://ostp.lib.iastate.edu/}}
  \label{fig:ResOrgs_allUS_vs_FFpercentage} 
\end{figure}

When looking by percentage, national laboratories again come to the top as expected. 
There is such tight clustering at the top that  data labels quickly become overlapped and unreadable; therefore a rectangular callout is added to Figure \ref{fig:ResOrgs_allUS_vs_FFpercentage} to show the grouping of national laboratories at high FF percentages.
The first non-federal agencies to appear are Scripps Research at \#21 and the Eli and Edythe L. Broad Institute of MIT and Harvard at \#22, both with almost exactly 73\% of their total output acknowledging federal funding.
Harvard University does appear one spot before that at \#20, but with their jointly administered Harvard - Smithsonian Center for Astrophysics.

Table \ref{tab:ResOrgs_topFFpercentage} shows the top 12 institutions with high FF percentages.

\begin{table}[h!tbp]
  \centering
  \caption{Research Institutions with highest percentage of FF articles}
    \begin{tabular}{|l|ccc|}
    Institution  & \multicolumn{1}{l}{FF Pubs} & \multicolumn{1}{l}{AllUS} & \multicolumn{1}{l}{Percentage} \\ \hline
    
    SLAC Nat Accel Lab &  3,861  &  4,411  & 87.53 \\
    Natl High Magnetic Field Lab & 1,345  & 1,555  & 86.50 \\
    Frederick Natl Lab for Cancer Research & 1,830  & 2,164  & 84.57 \\
    Lawrence Livermore Nat Lab & 6,082  & 7,232  & 84.10 \\
    Oak Ridge Nat Lab & 12,820  & 15,367  & 83.43 \\
    Argonne Nat Lab & 10,792  & 12,978  & 83.16 \\
    Brookhaven Nat Lab & 5,920  & 7,132  & 83.01 \\
    Natl Inst of Allergy and Infectious Diseases & 5,268  & 6,457  & 81.59 \\
    Pacific Northwest Nat Lab & 7,032  & 8,771  & 80.17 \\
    Lawrence Berkeley Nat Lab & 13,623  & 17,215  & 79.13 \\
    National Renewable Energy Lab & 4,099  &  5,185  & 79.05 \\
    Nal Center for Atmospheric Research & 2,944  & 3,733  & 78.86 \\
    
    \hline
    \end{tabular}%
  \label{tab:ResOrgs_topFFpercentage}%
\end{table}%

\subsection{RQ7 - Were these FF articles published Open Access or behind a paywall?}
\label{Results_subsec:RQ7}

So far, we have seen a detailed analysis of publisher, journal, and institution-level publication patterns, looking at how many U.S. federally funded articles were published over a certain time period out of a whole.
However, a question still remains - in what manner were those federally funded articles published?
Are they published as some form of Open Access and thus already freely available?
Or were they published behind a paywall, and if they had they been published in 2026 or later, would have represented a need to change the access mode?
In other words, how many of these past publications would have required a shift in access to comply with the Nelson OSTP memo guidance?

Open Access mode is provided to Dimensions by Unpaywall, an open database of free scholarly article metadata. 
Unpaywall determines the best OA location of a publication based on a cascading algorithm. It 
``prioritizes publisher-hosted content first (Hybrid or Gold), then prioritizes versions closer to the version of record (PublishedVersion over AcceptedVersion), then more authoritative repositories" \citep{Unpaywall_bestOAlocation}.
Therefore, even though a publication may match multiple OA codes, each publication receives only one OA status in Dimensions.
Dimensions also supplements Unpaywall's data with a list of full OA journals for the case of Gold Open Access \citep{Dimensions_OAcodes}.

An important nuance to keep in mind when looking at Open Access status is that Unpaywall provides the current status of an article as it appears when a report is run (in this case, as of October 2022).
There is a lack of historic OA data, so it is not possible to track the OA status at the time a publication appeared, or to find when an article qualified for a certain OA status.
Therefore, it is possible an article appearing in this analysis as a certain OA mode only earned that status recently and would have returned a different result if the analysis were run earlier.

\begin{figure}
    \centering
    \subfloat[\centering Federally funded publications]{{\includegraphics[width=0.47\textwidth]{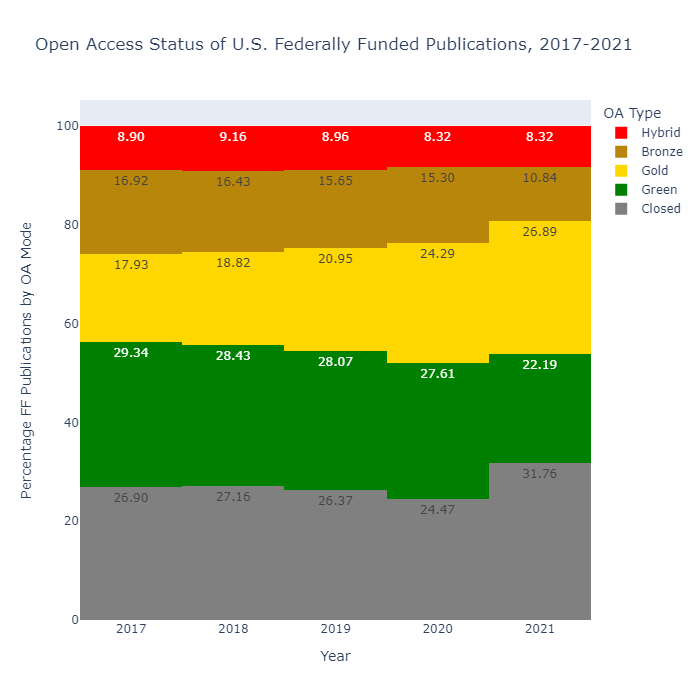}}}
    \qquad
    \subfloat[\centering All publications]{\includegraphics[width=0.47\textwidth]{{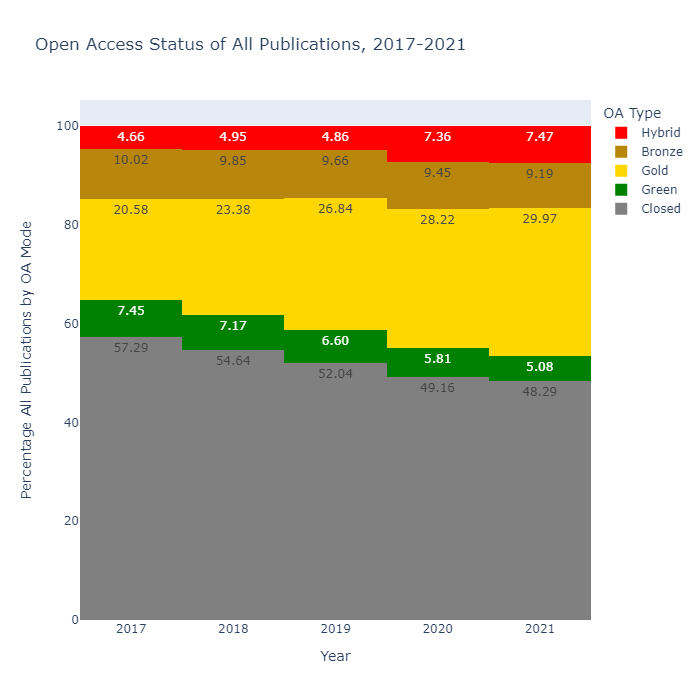}}}
    \caption{Open Access status of publications by year, 2017-2021}
    \label{fig:OAstatus_FF_and_all_subfigures}
\end{figure}

Figure \ref{fig:OAstatus_FF_and_all_subfigures}a displays the breakdown of each year's federally funded output, showing the percentage of FF articles published under each of five types of Open Access status.
For comparison, Figure \ref{fig:OAstatus_FF_and_all_subfigures}b shows the same plot, but for all publications worldwide from 2017-2021, roughly 5.6M per year.

Research outputs that were published Closed, or behind a paywall, appear at the bottom of each year's stack in gray. These are the publications that would have been affected the most by the new OSTP memo. Approximately 26\% of each year's FF papers were published Closed access over 2017-2020, but 2021 saw an increase to nearly 32\%, likely due to the 12-month embargo that is currently allowed by OSTP policy. Over time, the gray Closed access bar may return to a level more consistent with past years.

Green OA publications are self-archived by the author or a colleague by depositing the paper into a freely available university repository, disciplinary server, or to a personal webpage at no charge. This is the only OA route not delivered by the publisher, and the document may not exactly match the final version depending on publisher and journal restrictions.
Green OA has seen its share decrease over these years, from being the most common mode in 2017 to the third most common by 2021. 
This may again be an artifact of the currently allowed 12-month embargo period in the 2013 OSTP memo.
Authors may publish their work behind a paywall and make it publicly available through a Green OA route after 12 months.
Interestingly, compliance with the new OSTP memo could increase participation in Green OA, so this mode of access may dramatically increase once the guidance takes effect by 2026.

Gold OA refers to the final version of an article published in a fully OA journal that offers all articles immediately, permanently, and freely available on the journal website. This may or may not be the result of paying an article processing charge (APC).
Gold OA in federally funded publications has increased over the time period studied here, becoming the second most common mode of access in this dataset, at around 27\% in 2021.

Bronze OA is free access that is made temporarily available by the publisher, which can grant and remove access at any time without warning. It has seen the percentage of FF publications decrease over time.

Finally, Hybrid OA articles are published within a subscription (toll-based) journal, but made freely available on an individual, case-by-case basis by `unlocking" the article through paying an article processing charge (APC) to the publisher or journal. Hybrid contributes the smallest amount to the FF OA modes studied here, around 8\%.

In terms of all publications, \ref{fig:OAstatus_FF_and_all_subfigures}b shows a dramatic decrease in the number of Closed publications, with Gold increasing and taking a larger share each year.
Green remains relatively small at around 5\%, while Bronze and Hybrid make up 7 and 9\% yearly, respectively.

Papers that acknowledge U.S. federal funding are already much less likely to be published Closed access and much more likely to be deposited Green OA than a typical paper.
The impact of the OSTP memo will likely accelerate this, as depositing Green is one way to achieve zero-embargo availability.
Hybrid, Bronze, and Gold OA modes are roughly equivalent between U.S. federally funded and non- publications.

\subsubsection{Open Access by Publisher}
\begin{figure}[h!t] 
\centering 
  \includegraphics[width=1\textwidth]{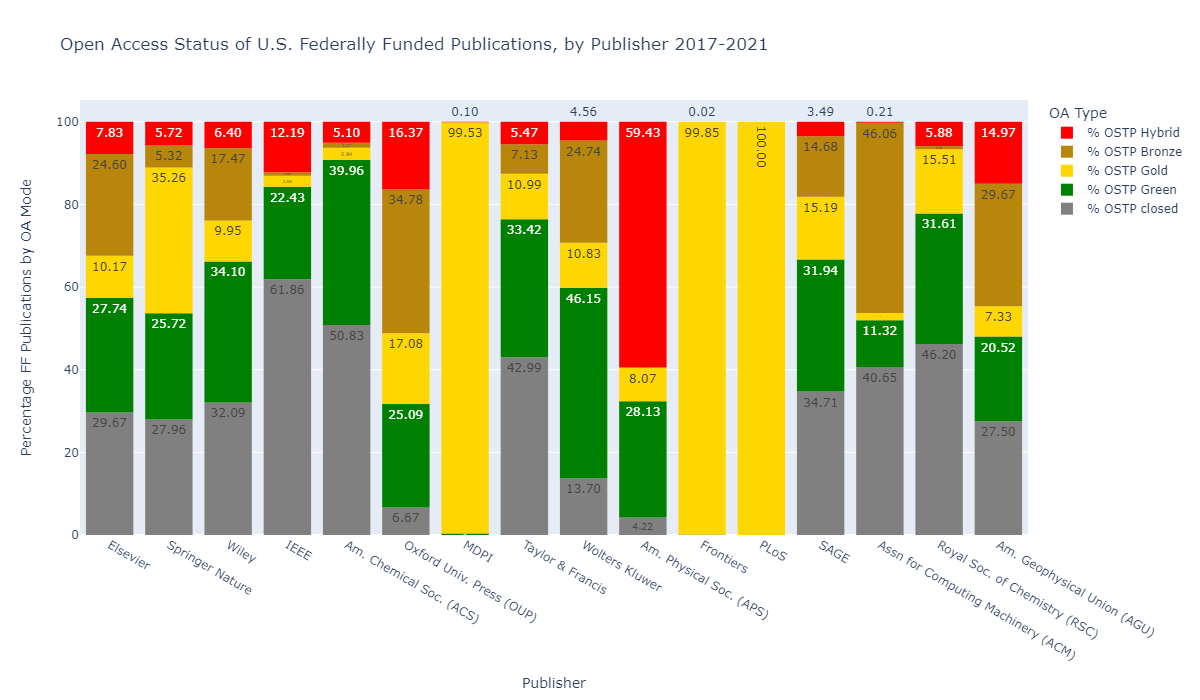} 
  \caption{Open Access status of FF Publications by Publisher, 2017-2021. Interactive version at \url{https://ostp.lib.iastate.edu/}}
  \label{fig:OA_FF_by_publisher_1-16} 
\end{figure}

Once OA status is introduced, we can combine some of the earlier aspects for further investigation.
Figure \ref{fig:OA_FF_by_publisher_1-16} shows a stacked barchart of the Open Access status of Federally Funded publications by publisher.
This shows the publishers with the 16 largest amounts of FF publications in terms of absolute number. 

Elsevier, Springer Nature, and Wiley are all large publishers with around 30\% of FF research published as Closed access.
IEEE and ACS have much higher percentages published as Closed, at around 50-60\%. 
These publishers may be more vulnerable to the change in policy by making their previously paywalled content publicly and openly available.
Pure Gold publishers appear strikingly as nearly 100\% yellow, such as MDPI, Frontiers, and PLoS.
Wolters Kluwer shows the highest amount of Green OA in this set of 16 publishers, at nearly 50\%.

This data could be presented in many other ways. Figure \ref{fig:OA_FF_by_publisher_1-16} shows the top 16 publishers by number of FF publications, but it may also be interesting to sort by highest percentage of Closed access or most Green OA.
A companion website is available at \url{https://ostp.lib.iastate.edu/} which expands the data presented in Figures \ref{fig:OA_FF_by_publisher_1-16} and \ref{fig:OA_FF_by_journal_1-16} to show the top 32 instead of only the top 16.
In addition, the user is able to change the sorting method. Choices include highest total number of FF publications or highest percentage of any of the OA status (Closed, Green, Gold, Bronze, or Hybrid).

\subsubsection{Open Access by Journal Title}

\begin{figure}[h!t] 
\centering 
  \includegraphics[width=1\textwidth]{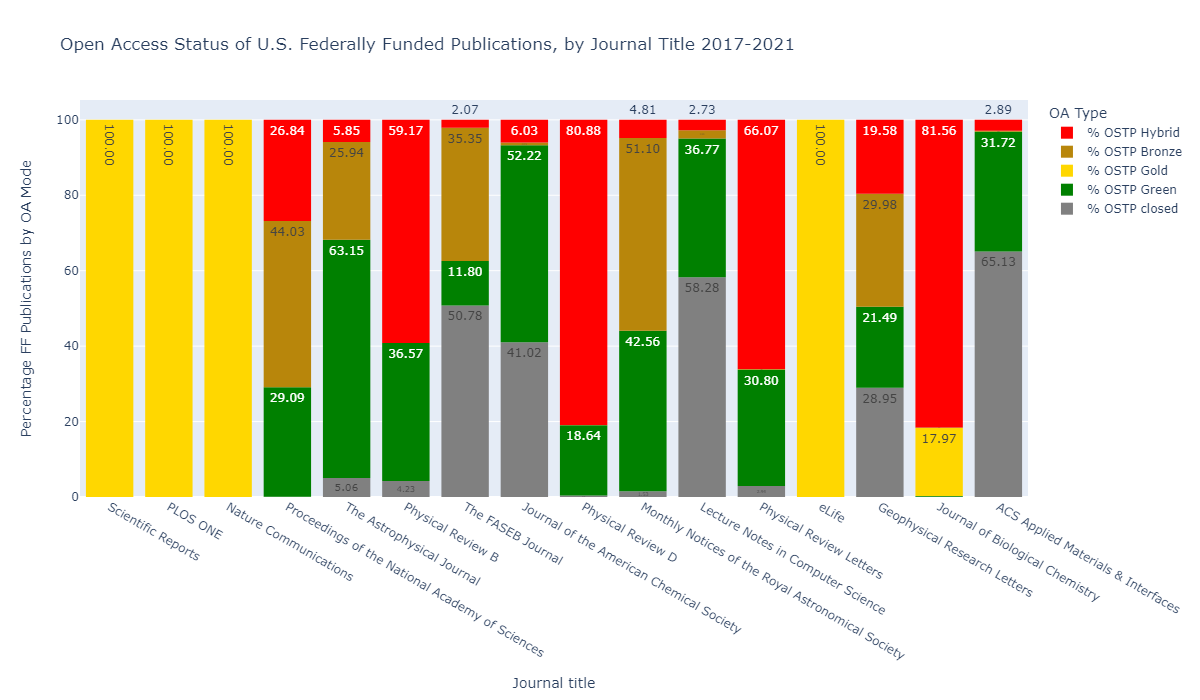} 
  \caption{Open Access status of FF Publications by Journal Title, 2017-2021. Interactive version at \url{https://ostp.lib.iastate.edu/}}
  \label{fig:OA_FF_by_journal_1-16} 
\end{figure}

Moving to the OA status of individual journal titles, we can again see certain journals have the potential to be affected more heavily than others.
Figure \ref{fig:OA_FF_by_journal_1-16} shows the top 16 journal titles by absolute number of FF publications over the five years studied.
The top three journal titles are all 100\% Gold OA: Scientific Reports, PLOS ONE, and Nature Communications. eLife also appears as completely Gold, at number 13. Presumably, these journals could continue operating as they are today even after the new OSTP policy framework takes effect in 2026. 
It is possible there is already some portion of FF output in these journals that is deposited Green OA. Unpaywall's algorithm prefers publisher-provided Open Access over repository provided OA, so the presence of Green OA within these journals would require a deeper analysis beyond what Unpaywall reports as the ``best OA status."

Conversely, the FASEB Journal, Journal of the ACS, Lecture Notes in Computer Science, and ACS Applied Materials \& Interfaces all publish a substantial proportion of their total federally funded output behind a paywall.
These journals will need to adjust their policies and strategy in order to comply with the coming guidance of making FF publications immediately and publicly available.

Similar to publisher OA data, the companion website also allows a user to sort journal title OA data by highest total number of FF publications, or the highest percentage of any OA mode.
\section{Limitations}
\label{sec:Limitations}
The clearest limitation of this analysis is the  likelihood that not all U.S. federally funded research is included in the dataset.
This study is limited by the fact that it only knows about publications that identify funding sources, and there is a possibility that some were missed.
For example, the NSF's Polar Environment, Safety and Health Section (PESH) is a valid granting agency, but it returns 0 results when directly searched for in Dimensions.
PESH is included in the list of agencies that make up the custom filter but it has no effect on the number of publications analyzed here.

As noted in Section \ref{sec:Methodology}, the availability of funding information from Crossref varies widely by publisher.
This becomes the starting point for Dimensions, who then enriches the information with full text analysis and agreements with publishers to to obtain additional funding information.
Even when it is included, not all publications correctly name the funder or provide a funder ID. 
The new OSTP memo addresses this in section 4.a:
``Agencies should...collect and make publicly available appropriate metadata associated with scholarly publications and data resulting from federally funded research...Such metadata should include at minimum: all author and co-author names, affiliations, and sources of funding, referencing digital persistent identifiers, as appropriate" \citep{Nelson_memo}.

Authors will need to accurately and appropriately report their funding information, publishers need to supply that information to Crossref when registering for a DOI, and bibliographic databases must to ingest that information.
Enriching and enhancing funding information by analyzing the text of the acknowledgments section is also helpful, but could be improved.

It is also possible that some federal funding agencies were not explicitly included in the custom-made filter. 
Three rounds of refinement captured 239 individual funding agencies.
One could always go further, but at some point, there are diminishing returns to continuing to export funders, manually assess them one by one, and add them to the custom Dimensions filter.
A researcher with more in-depth knowledge of U.S. federal grant funding agencies and their sub-divisions, institutes, and centers could investigate the custom filter that was defined and used here to identify holes or gaps.

The document type Article in Dimensions covers many  types of content in journals, including editorials, letters, corrections, book reviews, news items, etc
\citep{Dimensions_whats_in_Article}.
These materials were unable to be separated from the main Article type and were therefore included in this analysis.

\section{Conclusion}
The practical implications of the August 2022 OSTP memo's guidance are still being defined.
Making federally funded publications immediately publicly available will involve a shift in strategy and behavior for publishers, authors, institutions, and readers.
These peer-reviewed publications becoming immediately accessible to the public will expand the level of impact and reach, but it may also bring with it some ramifications that may not yet not be completely understood.

Though the OSTP released a companion impact report, it did not investigate the potential effects beyond a general estimation of the number of articles affected per year.
This analysis went further but is only a first step in attempting to understand the broad-reaching implications of this updated policy.
Quantifying the number, nature, and characteristics of publications from the past which would have qualified under this policy framework helps to clarify some questions and provide guidance still outstanding.
It is clear that publishers, journals, and research institutions will all be affected, with some needing to adjust more than others.
Once the new OSTP guidance takes effect, the equivalent of Figures \ref{fig:OAstatus_FF_and_all_subfigures}a, \ref{fig:OA_FF_by_publisher_1-16}, and
\ref{fig:OA_FF_by_journal_1-16} will all become completely Green, Gold, or Hybrid.

Reported funder information is critical and will remain important as the OSTP guidance takes effect in 2026. 
Publishers, funders, and authors need to submit complete, accurate, and structured funding information, and database providers should continue to extract additional information to enhance this metadata.
Dimensions and other bibliographic metadata tools will continue to define and refine funder filters to enable users to conduct a similar analysis to this on their own institution's publications.

\subsection{Acknowledgements}
This paper was written using data obtained on October 18, 2022 from the paid version of \citeauthor{Dimensions_itself}'s Dimensions platform, available at \url{https://app.dimensions.ai}. 
Plots were created using Plotly version 5.10.0 \citep{Plotly_itself}.
The Dimensions Support Team was also very helpful in answering questions related to the construction of custom groups and providing further details on the data.

\subsection{Competing Interests}
The author declares no competing interests.

\subsection{Funding Information}
The author declares no funding was received.

\subsection{Data Availability}
The data resulting from this research are available in .csv format at \url{https://doi.org/10.5281/zenodo.7254815}
and
\url{https://github.com/eschares/OSTP_impact} \citep{Schares_OSTP_Impact}.

\begin{itemize}
    \item Data set of funders
    \item Data set of publishers
    \item Data set of journal titles
    \item Data set of research organizations
    \item Data set of Open Access status by FF and worldwide
\end{itemize}

A companion website is also available at \url{https://ostp.lib.iastate.edu/} which includes interactive versions of each plot shown in this paper. 
Users may pan, zoom, and hover over data points for more information.
Additionally, they may search for a specific publisher, journal title, or research institution to enable its data label and color it red for easier identification on the graph.

\bibliography{bib.bib}

\newpage
\begin{center}
    \section*{Appendix}
    \label{sec:Appendix}
\end{center}

\begin{longtable}{| p{.20\textwidth} | p{.80\textwidth} |} 
\caption{237 U.S. federal funding agencies included in custom filter}\\
    Name  & ID \\
    \hline
    ID    & Name \\
    grid.473856.b & Administration for Children and Families (ACF) \\
    grid.473794.a & Administration for Community Living (ACL) \\
    grid.452934.d & Advanced Research Projects Agency-Energy (ARPA-E) \\
    grid.413404.6 & Agency for Healthcare Research and Quality (AHRQ) \\
    grid.453168.d & Agency for Toxic Substances and Disease Registry (ATSDR) \\
    grid.463419.d & Agricultural Research Service (ARS) \\
    grid.427848.5 & Air Force Institute of Technology (AFIT) \\
    grid.419075.e & Ames Research Center (ARC) \\
    grid.413759.d & Animal and Plant Health Inspection Service (APHIS) \\
    grid.187073.a & Argonne National Laboratory (ANL) \\
    grid.202665.5 & Brookhaven National Laboratory (BNL) \\
    grid.468484.2 & Center for Enabling New Technologies Through Catalysis (CENTC) \\
    grid.410422.1 & Center for Information Technology (CIT) \\
    grid.473771.1 & Center for Neuroscience and Regenerative Medicine (CNRM) \\
    grid.453876.b & Center for Scientific Review (CSR) \\
    grid.457888.e & Center for Selective C–H Functionalization (NSF CCHF) \\
    grid.416738.f & Centers for Disease Control and Prevention (CDC) \\
    grid.413874.d & Centers for Medicare and Medicaid Services (CMS) \\
    grid.453107.4 & Climate Program Office (CPO) \\
    funder.201191 & Combat Casualty Care Research Program (CCCRP) \\
    grid.496791.4 & Congressionally Directed Medical Research Programs (CDMRP) \\
    grid.421816.8 & Defense Advanced Research Projects Agency (DARPA) \\
    grid.453336.4 & Defense Logistics Agency (DLA) \\
    grid.452918.3 & Defense Threat Reduction Agency (DTRA) \\
    funder.57700 & Department of Defense, Small Business Innovation Research (DOD SBIR) \\
    grid.457768.f & Directorate for Biological Sciences (NSF BIO) \\
    grid.457785.c & Directorate for Computer \& Information Science \& Engineering (NSF CISE) \\
    grid.457799.1 & Directorate for Education \& Human Resources (NSF EHR) \\
    grid.457810.f & Directorate for Engineering (NSF ENG) \\
    grid.457836.b & Directorate for Geosciences (NSF GEO) \\
    grid.457875.c & Directorate for Mathematical \& Physical Sciences (NSF MPS) \\
    grid.457916.8 & Directorate for Social, Behavioral \& Economic Sciences (NSF SBE) \\
    grid.515271.5 & Directorate for Technology, Innovation and Partnerships (NSF TIP) \\
    grid.467627.6 & Division of Acquisition and Cooperative Support (DACS) \\
    grid.467624.5 & Division of Antarctic Infrastructure and Logistics (NSF AIL) \\
    grid.457878.1 & Division of Astronomical Sciences (NSF AST) \\
    grid.457839.4 & Division of Atmospheric and Geospace Sciences (NSF AGS) \\
    grid.457920.d & Division of Behavioral and Cognitive Sciences (NSF BCS) \\
    grid.457770.6 & Division of Biological Infrastructure (NSF DBI) \\
    grid.457813.c & Division of Chemical, Bioengineering, Environmental, and Transport Systems (NSF CBET) \\
    grid.457885.3 & Division of Chemistry (NSF CHE) \\
    grid.457814.b & Division of Civil, Mechanical \& Manufacturing Innovation (NSF CMMI) \\
    grid.457794.c & Division of Computer and Network Systems (NSF CNS) \\
    grid.457793.b & Division of Computing and Communication Foundations (NSF CCF) \\
    grid.457842.8 & Division of Earth Sciences (NSF EAR) \\
    grid.457818.7 & Division of Electrical, Communications \& Cyber Systems (NSF ECCS) \\
    grid.467634.4 & Division of Elementary, Secondary, and Informal Education (ESIE) \\
    grid.457821.d & Division of Engineering Education \& Centers (NSF EEC) \\
    grid.457772.4 & Division of Environmental Biology (NSF DEB) \\
    grid.457801.f & Division of Graduate Education (NSF DGE) \\
    grid.457762.5 & Division of Grants \& Agreements (NSF DGA) \\
    grid.457805.b & Division of Human Resource Development (NSF HRD) \\
    grid.467631.1 & Division of Human Resource Management (NSF HRM) \\
    grid.457832.f & Division of Industrial Innovation \& Partnerships (NSF IIP) \\
    grid.457797.f & Division of Information and Intelligent Systems (NSF IIS) \\
    grid.457911.f & Division of Information Systems (NSF DIS) \\
    grid.467630.0 & Division of Institution and Award Support (DIAS) \\
    grid.457779.f & Division of Integrative Organismal Systems (NSF IOS) \\
    grid.457891.6 & Division of Materials Research (NSF DMR) \\
    grid.457892.5 & Division of Mathematical Sciences (NSF DMS) \\
    grid.457783.a & Division of Molecular \& Cellular Biosciences (NSF MCB) \\
    grid.457845.f & Division of Ocean Sciences (NSF OCE) \\
    grid.457893.4 & Division of Physics (NSF PHY) \\
    grid.454854.c & Division of Program Coordination Planning and Strategic Initiatives (DPCPSI) \\
    grid.457802.c & Division of Research on Learning in Formal and Informal Settings (NSF DRL) \\
    grid.457922.f & Division of Social and Economic Sciences (NSF SES) \\
    grid.457803.d & Division of Undergraduate Education (NSF DUE) \\
    grid.482913.5 & Economic Research Service (ERS) \\
    grid.457773.5 & Emerging Frontiers Office (NSF EF) \\
    grid.417553.1 & Engineer Research and Development Center (ERDC) \\
    grid.418698.a & Environmental Protection Agency (EPA) \\
    grid.420089.7 & Eunice Kennedy Shriver National Institute of Child Health and Human Development (NICHD) \\
    grid.421881.6 & Federal Emergency Management Agency (FEMA) \\
    grid.483785.5 & Federal Highway Administration (FHWA) \\
    grid.417851.e & Fermilab (Fermilab) \\
    grid.453035.4 & Fogarty International Center (FIC) \\
    grid.419077.c & Glenn Research Center (Glenn Research Center) \\
    grid.133275.1 & Goddard Space Flight Center (GSFC) \\
    grid.454842.b & Health Resources and Services Administration (HRSA) \\
    grid.483884.b & Hydrogen and Fuel Cells Technologies Office (HFTO) \\
    grid.414598.5 & Indian Health Service (IHS) \\
    grid.421530.1 & Institute of Education Sciences (IES) \\
    grid.431754.0 & Institute of Museum and Library Services (IMLS) \\
    grid.211367.0 & Jet Propulsion Lab (JPL) \\
    grid.419085.1 & Johnson Space Center (JSC) \\
    grid.419743.c & Kennedy Space Center (KSC) \\
    grid.419086.2 & Langley Research Center (LaRC) \\
    grid.184769.5 & Lawrence Berkeley National Laboratory (LBL) \\
    grid.250008.f & Lawrence Livermore National Laboratory (LLL) \\
    grid.148313.c & Los Alamos National Laboratory (LANL) \\
    grid.419091.4 & Marshall Space Flight Center (MSFC) \\
    grid.463240.7 & Missile Defense Agency (MDA) \\
    grid.238252.c & National Aeronautics and Space Administration (NASA) \\
    grid.48336.3a & National Cancer Institute (NCI) \\
    grid.429651.d & National Center for Advancing Translational Sciences (NCATS) \\
    grid.57828.30 & National Center for Atmospheric Research (NCAR) \\
    grid.416781.d & National Center for Chronic Disease Prevention and Health Promotion (NCCDPHP) \\
    grid.280655.c & National Center for Complementary and Integrative Health (NCCIH) \\
    grid.467923.d & National Center for Emerging and Zoonotic Infectious Diseases (NCEZID) \\
    grid.416778.b & National Center for Environmental Health (NCEH) \\
    grid.416789.5 & National Center for Health Statistics (NCHS) \\
    grid.419980.d & National Center for HIV/AIDS Viral Hepatitis STD and TB Prevention (NCHSTP) \\
    grid.419260.8 & National Center for Immunization and Respiratory Diseases (NIP) \\
    grid.453275.2 & National Center for Injury Prevention and Control (NCIPC) \\
    grid.457921.c & National Center for Science and Engineering Statistics (NSF NCSE) \\
    grid.453445.7 & National Center on Birth Defects and Developmental Disabilities (NCBDD) \\
    grid.423033.5 & National Centers for Coastal Ocean Science (NCCOS) \\
    grid.422239.c & National Endowment for the Humanities (NEH) \\
    grid.451363.6 & National Energy Technology Laboratory (NETL) \\
    grid.280030.9 & National Eye Institute (NEI) \\
    grid.462540.1 & National Geospatial-Intelligence Agency (NIMA) \\
    grid.279885.9 & National Heart Lung and Blood Institute (NHLBI) \\
    grid.467654.2 & National Highway Traffic Safety Administration (NHTSA) \\
    grid.280128.1 & National Human Genome Research Institute (NHGRI) \\
    grid.451056.3 & National Institute for Health Research (NIHR) \\
    grid.457946.d & National Institute for Mathematical and Biological Synthesis (NIMBioS) \\
    grid.416809.2 & National Institute for Occupational Safety and Health (NIOSH) \\
    grid.419681.3 & National Institute of Allergy and Infectious Diseases (NIAID) \\
    grid.420086.8 & National Institute of Arthritis and Musculoskeletal and Skin Diseases (NIAMS) \\
    grid.280347.a & National Institute of Biomedical Imaging and Bioengineering (NIBIB) \\
    grid.419633.a & National Institute of Dental and Craniofacial Research (NIDCR) \\
    grid.419635.c & National Institute of Diabetes and Digestive and Kidney Diseases (NIDDK) \\
    grid.280664.e & National Institute of Environmental Health Sciences (NIEHS) \\
    grid.482914.2 & National Institute of Food and Agriculture (NIFA) \\
    grid.280785.0 & National Institute of General Medical Sciences (NIGMS) \\
    grid.484220.8 & National Institute of Justice (NIJ) \\
    grid.416868.5 & National Institute of Mental Health (NIMH) \\
    grid.416870.c & National Institute of Neurological Disorders and Stroke (NINDS) \\
    grid.280738.6 & National Institute of Nursing Research (NINR) \\
    grid.94225.38 & National Institute of Standards and Technology (NIST) \\
    grid.419475.a & National Institute on Aging (NIA) \\
    grid.420085.b & National Institute on Alcohol Abuse and Alcoholism (NIAAA) \\
    grid.214431.1 & National Institute on Deafness and Other Communication Disorders (NIDCD) \\
    grid.473857.a & National Institute on Disability, Independent Living, and Rehabilitation Research (NIDILRR) \\
    grid.420090.f & National Institute on Drug Abuse (NIDA) \\
    grid.281076.a & National Institute on Minority Health and Health Disparities (NIMHD) \\
    grid.94365.3d & National Institutes of Health (NIH) \\
    grid.410305.3 & National Institutes of Health Clinical Center (CLC) \\
    grid.453023.3 & National Nuclear Security Administration (NNSA) \\
    grid.3532.7 & National Oceanic and Atmospheric Administration (NOAA) \\
    grid.454846.f & National Park Service (NPS) \\
    grid.454847.e & National Pork Board \\
    grid.419357.d & National Renewable Energy Laboratory (NREL) \\
    grid.457896.1 & National Science Board (NSF NSB) \\
    grid.431093.c & National Science Foundation (NSF) \\
    grid.482831.4 & National Security Agency (NSA) \\
    grid.454848.1 & National Space Biomedical Research Institute (NSBRI) \\
    grid.420427.7 & Naval Air Systems Command (NAVAIR) \\
    grid.419445.9 & Naval Information Warfare Center Pacific (SPAWAR) \\
    grid.415913.b & Naval Medical Research Center (NMRC) \\
    grid.462643.1 & Naval Sea Systems Command (NAVSEA) \\
    grid.473834.f & NIHR Academy (TCC) \\
    grid.473755.7 & NIHR Central Commissioning Facility (CCF) \\
    grid.473757.5 & NIHR Evaluation Trials and Studies Coordinating Centre (NETS) \\
    grid.135519.a & Oak Ridge National Laboratory (ORNL) \\
    grid.467641.6 & Office for State, Tribal, Local and Territorial Support (OSTLTS) \\
    grid.457789.0 & Office of Advanced Cyberinfrastructure (NSF OAC) \\
    grid.453070.5 & Office of Advanced Scientific Computing Research (ASCR) \\
    grid.452988.a & Office of Basic Energy Sciences (BES) \\
    grid.452963.f & Office of Biological and Environmental Research (BER) \\
    grid.457758.c & Office of Budget, Finance and Award Management (NSF BFA) \\
    grid.453518.e & Office of Dietary Supplements (ODS) \\
    grid.453355.1 & Office of Electricity Delivery and Energy Reliability (OE) \\
    grid.457829.5 & Office of Emerging Frontiers and Multidisciplinary Activities (NSF EFMA) \\
    grid.453003.1 & Office of Energy Efficiency and Renewable Energy (EERE) \\
    grid.457407.5 & Office of Environmental Management (EM) \\
    grid.436341.7 & Office of Extramural Research (OER) \\
    grid.453259.c & Office of Fossil Energy (FE) \\
    grid.453030.1 & Office of Fusion Energy Sciences (FES) \\
    grid.457907.8 & Office of Information and Resource Management (NSF OIRM) \\
    grid.473792.c & Office of Inspector General (OIG) \\
    grid.457900.f & Office of Integrative Activities (NSF OIA) \\
    grid.457904.b & Office of International Science and Engineering (NSF OISE) \\
    grid.433602.0 & Office of Juvenile Justice and Delinquency Prevention (OJJDP) \\
    grid.467629.8 & Office of Legislative \& Public Affairs (OLPA) \\
    grid.467644.3 & Office of Multidisciplinary Activities (OMA) \\
    grid.482851.2 & Office of Naval Research (ONR) \\
    grid.453031.0 & Office of Nuclear Energy (NE) \\
    grid.453025.5 & Office of Nuclear Physics (NP) \\
    grid.457846.c & Office of Polar Programs (NSF PLR) \\
    grid.467635.5 & Office of Public Health Preparedness and Response (OPHPR) \\
    grid.453216.7 & Office of Science (DOE SC) \\
    grid.483892.c & Office of Special Education and Rehabilitative Services (OSERS) \\
    grid.457898.f & Office of the Director (NSF OD) \\
    grid.453125.4 & Office of the Director (OD) \\
    grid.488591.9 & Office of the Director of National Intelligence (ODNI) \\
    grid.473761.0 & Office of the Secretary of Defense (OSD) \\
    grid.467626.7 & Polar Environment, Safety and Health Section (PESH) \\
    grid.458401.c & President's Emergency Plan for AIDS Relief (PEPFAR) \\
    grid.458379.4 & Quality Enhancement Research Initiative (QUERI) \\
    grid.453134.4 & Rehabilitation Research and Development Service (JRRD) \\
    grid.467664.1 & Savannah River Operations Office (DOE-SR) \\
    grid.457927.a & SBE Office of Multidisciplinary Activities (NSF SMA) \\
    grid.457924.9 & Science of Learning Centers Program Coordinating Committee (NSF SLC-CC) \\
    grid.453092.9 & Small Business Administration (SBA) \\
    grid.413730.2 & Substance Abuse and Mental Health Services Administration (SAMHSA) \\
    grid.453220.2 & Telemedicine \& Advanced Technology Research Center (TATRC) \\
    grid.513044.7 & Traumatic Brain Injury Center of Excellence (TBICoE) \\
    grid.265436.0 & Uniformed Services University of the Health Sciences (USUHS) \\
    grid.420285.9 & United States Agency for International Development (USAID) \\
    grid.453002.0 & United States Air Force (USAF) \\
    grid.265457.7 & United States Air Force Academy (USAFA) \\
    grid.507554.6 & United States Air Force Office of Scientific Research (AFOSR) \\
    grid.417730.6 & United States Air Force Research Laboratory (AFRL) \\
    grid.420176.6 & United States Army (USA) \\
    grid.431335.3 & United States Army Corps of Engineers (CoE) \\
    grid.481489.8 & United States Army Medical Command (MEDCOM) \\
    grid.420210.5 & United States Army Medical Research and Development Command (USAMRMC) \\
    grid.416900.a & United States Army Medical Research Institute of Infectious Diseases (USAMRIID) \\
    grid.420282.e & United States Army Research Laboratory (ARL) \\
    grid.507553.1 & United States Army Research Office (US ARO) \\
    grid.418402.b & United States Army Research, Development and Engineering Command (RDECOM) \\
    grid.417548.b & United States Department of Agriculture (USDA) \\
    grid.239119.1 & United States Department of Commerce (USDoC) \\
    grid.420391.d & United States Department of Defense (USDOD ) \\
    grid.419881.d & United States Department of Education (DoED) \\
    grid.85084.31 & United States Department of Energy (DOE) \\
    grid.27235.31 & United States Department of Health and Human Services (DHHS) \\
    grid.433574.2 & United States Department of Homeland Security (DHS) \\
    grid.420410.3 & United States Department of Housing and Urban Development (HUD) \\
    grid.423379.8 & United States Department of Justice (DOJ) \\
    grid.495456.f & United States Department of the Air Force (DAF) \\
    grid.427904.c & United States Department of the Army (DA) \\
    grid.239134.e & United States Department of the Interior (DOI) \\
    grid.420434.5 & United States Department of the Navy (DON) \\
    grid.419882.e & United States Department of Transportation (USDOT) \\
    grid.418356.d & United States Department of Veterans Affairs (VA ) \\
    grid.462979.7 & United States Fish and Wildlife Service (FWS) \\
    grid.417587.8 & United States Food and Drug Administration (FDA ) \\
    grid.2865.9 & United States Geological Survey (USGS) \\
    grid.497574.c & United States Marine Corps (USMC) \\
    grid.280285.5 & United States National Library of Medicine (NLM) \\
    grid.431418.b & United States Nuclear Regulatory Commission (NRC) \\
    grid.417684.8 & United States Public Health Service (USPHS) \\
    grid.472551.0 & US Forest Service (USFS) \\
    grid.483017.9 & USDA Rural Development (USDA RD) \\
    grid.453027.7 & Vehicle Technologies Office (Vehicle Technologies Office) \\
    grid.507680.c & Walter Reed Army Institute of Research (WRAIR) \\
    grid.453840.e & Women's Health Initiative (WHI) \\
    \hline
\end{longtable}
\label{tab:FullFunderList}

\end{document}